\documentclass[10pt,journal]{IEEEtran}
\IEEEoverridecommandlockouts
\usepackage{mathrsfs}
\usepackage{amssymb}
\usepackage{cite}
\UseRawInputEncoding

\ifCLASSINFOpdf
  \usepackage[pdftex]{graphicx}
     \graphicspath{{../pdf/}{../jpeg/}}
    \DeclareGraphicsExtensions{.pdf,.jpeg,.png}
\else
   \usepackage[dvips]{graphicx}
   \graphicspath{{../eps/}}
   \DeclareGraphicsExtensions{.eps}
\fi
\usepackage[cmex10]{amsmath}
\usepackage{algorithmicx}
\usepackage{algpseudocode}
\usepackage{algorithm}
\usepackage{array}
\usepackage{bm}
\usepackage{eqparbox}
\usepackage{mdwmath}
\usepackage{mdwtab}
\usepackage[tight,footnotesize]{subfigure}
\usepackage[caption=1]{caption}
\usepackage[font=footnotesize]{subfig}
\usepackage{fixltx2e}

\usepackage{url}

\begin{document}
\title {Passive Beamforming and Fronthaul Compression Design for Intelligent Reflecting Surface-aided Cloud Radio Access Networks}
\author{Yu~Zhang, Xuelu Wu, Hong Peng, Caijun Zhong and Xiaoming Chen
\thanks{} 
\thanks{

Y. Zhang, X. Wu and H. Peng are with the College of Information Engineering, Zhejiang University of Technology, China. (e-mail:{\tt yzhang@zjut.edu.cn}, {\tt ph@zjut.edu.cn}). Y. Zhang is also with the National Mobile Communications Research Laboratory, Southeast University, China.

C. Zhong ang X. Chen are with the College of Information Science and Electronic Engineering, Zhejiang University, China (e-mail: {\tt caijunzhong@zju.edu.cn}, {\tt chen\_xiaoming@zju.edu.cn})

%
}
\thanks{}} \maketitle

\begin{abstract}
This letter studies a cloud radio access network (C-RAN) with multiple intelligent reflecting surfaces (IRS) deployed between users and remote radio heads (RRH). Specifically, we consider the uplink transmission where the RRHs quantize the received signals from the users by either point-to-point compression or Wyner-Ziv coding and then transmits the quantization bits to the BBU pool through capacity limited fronthaul links. To maximize the uplink sum rate, we jointly optimize the IRS passive beamformers and the quantization noise covariance matrices for fronthaul compression. By exploiting the Arimoto-Blahut algorithm and semidefinite relaxation, we propose a successive approximation and optimization approach to solve the non-convex problem. Numerical results verify the performance gain achieved by the proposed design.
\end{abstract}
\begin{IEEEkeywords}
C-RAN, IRS, fronthaul compression, Arimoto-Blahut algorithm.
\end{IEEEkeywords}
\section{Introduction}
Cloud radio access network (C-RAN) is a prospective mobile network architecture, which provides an efficient way for multi-cell interference management \cite{Peng2015CRAN}. In a C-RAN, the baseband processing function of conventional base stations is backward migrated into a baseband unit (BBU) pool and radio remote heads (RRH) are deployed closer to users. Nevertheless, high-speed fronthaul links are required to connect the RRHs and BBU pool, which restricts the flexible and dense deployment of RRHs due to the implementation cost and complexity in practice.

To tackle this issue, we propose to exploit the recently-emerging intelligent reflecting surface (IRS) to enhance the access link between users and RRHs in the C-RAN. IRS consists of a large number of reflecting elements with which controllable phase shifts can impose on the imping waves \cite{IRS-asid-wireless}. Benefited from this, IRS can generate desired reflection beams and create favorable propagation conditions. Since IRS is basically a passive device and solely requires a low-rate control link, it provides an energy-efficient and cost-effective way to enhance the C-RAN. Most recently, a plethora of works have studied the design of IRS-assisted wireless communication systems\cite{Programble}-\cite{Statiscal-CSI}.
In particular, IRS was considered in multi-cell systems to assist coordinated multiple point transmission\cite{multipoint}. The IRS-aided cell-free systems were investigated in \cite{Cell-free1}, where the weighted system sum rate was maximized. Nevertheless, backhaul link capacity limit was not considered in the above works. For C-RAN, we considered to deploy a single IRS on the wireless fronthaul to enhance the uplink rate\cite{wirelessfronthual}. The authors in \cite{Over-the-Air} exploited IRS to improve the accuracy of the over-the-air computation in C-RAN.

In this letter, we focus on the uplink transmission design for a C-RAN wherein multiple IRSs are deployed on the access link between users and RRHs. Due to the limited fronthaul capacity, the received signals at each RRH are compressed first before being conveyed to the BBU pool, using point-to-point compression or Wyner-Ziv coding\cite{Park14fronth}.
To fully exploit the advantage of IRS, the IRS passive beamforming should be jointly designed together with the fronthaul compression, which has not been considered in the aforementioned works for multi-cell systems and cell-free networks. In this letter, firstly, with the goal of maximizing the uplink sum rate, we jointly optimize the IRS beamformers and the quantization noise covariance matrices under Wyner-Ziv coding. In particular, we consider Wyner-Ziv coding with joint decompression instead of sequential decompression as in \cite{wirelessfronthual} and \cite{Zhou2016CRANopt}, which is more general and leads to different fronthaul constraints. By exploiting the Arimoto-Blahut algorithm \cite{Two-way} and semi-definite relaxation (SDR), we propose a successive approximation and optimization algorithm to solve the non-convex problem. Then we show that the proposed algorithm can be simply extended to the case of point-to-point compression. Moreover, inspired by the results in conventional C-RAN \cite{Zhou2016CRANopt}, we modify the proposed algorithm for high signal-to-noise-ratio (SQNR) regime, with which the computational complexity is further reduced. Finally, numerical results show the performance gain of deploying IRS achieved by the proposed algorithms.

\textit{Notation}: For a matrix $\bf A$, $|{\bf A}|$, $\mathrm{Tr}\left({\bf A}\right)$, ${\bf A}^T$ and ${\bf A}^H$ denote the determinant, trace, transpose and conjugate transpose of $\bf A$. $\mathrm{{diag}}\left(\bf A\right)$ denotes a column vector formed with the diagonals of $\bf A$. For an index set $\mathcal S$, unless otherwise specified, ${\bf A}_{\mathcal S}$ denotes the matrix with elements ${\mathbf{A}}_{i}$ whose indices $i \in {\mathcal S}$ and ${\rm{diag}}\left( {{{\left\{ {{{\bf{A}}_i}} \right\}}_{i \in {\mathcal S}}}} \right)$ denotes the block diagonal matrix formed with ${\bf A}_{i}$ on the diagonal where $i \in {\mathcal S}$. ${\bf A}\odot{\bf B}$ denotes the Hadamard product of $\bf A$ and $\bf B$. ${\bf I}$ denotes the identity matrix. $\mathbb{E}\left[.\right]$ stands for the expectation. Let $\mathcal{K}=\{1,\cdots,K\}$, $\mathcal{L}=\{1,\cdots,L\}$ and $\mathcal{M}=\{1,\cdots,M\}$.

\section{prelimilary}
\subsection{System Model}
We consider the uplink transmission of a C-RAN, where \emph{K} single-antenna users communicate with the BBU pool through \emph{L} RRHs, each equipped with ${N_R}$ antennas. \emph{M} IRSs are deployed to aid the communication between users and RRHs, each of which has ${N_I}$ reflecting elements. For simplicity, we assume global channel state information (CSI) at the BBU pool. Note that the CSI acquisition for the IRS link has been discussed, e.g., in \cite{CSI-acquisition}.

On the access link, user $k\in\mathcal{K}$ transmits the signal ${x_k}$ to RRHs. Let $\mathbf{x} = {[{x_1},...,{x_k}]^T}$ with $\mathbf{x} \sim \mathcal{CN}(\mathbf{0},P\bf{I})$, where $P$ denotes user transmit power. Then the signals received by RRH $l\in\mathcal{L}$ can be expressed as
\begin{equation}\label{trans_model}
\begin{split}
{\mathbf{y}_l} &= {\mathbf{H}_{l}}\mathbf{x} + \sum\limits_{m = 1}^M {{\mathbf{G}_{l,m}}{\mathbf{\Theta} _m}} {\mathbf{H}_{R,m}}\mathbf{x} + {\mathbf{n}_l}\\
&= ({\mathbf{H}_{l}} + {\mathbf{G}_{l,\mathcal{M}}}\mathbf{\Theta} {\mathbf{H}_{R,\mathcal{M}}})\mathbf{x} + {\mathbf{n}_l},
\end{split}
\end{equation}
where ${{\bf{H}}_l} \in {\mathbb{C}^{{N_R} \times K}}$, ${{\bf{G}}_{l,m}} \in {\mathbb{C}^{{N_R} \times {N_I}}}$ and ${{\bf{H}}_{R,m}} \in {\mathbb{C}^{{N_I} \times K}}$ represent the channel matrix between users and RRH \emph{l}, between IRS \emph{m} and RRH \emph{l}, and between users and IRS \emph{m}, respectively, ${\mathbf{\Theta} _m} = \mathrm{diag}({\theta _{m,1}},...,{\theta _{m,N_I}})$ represents the passive beamformer of IRS $m$ (we assume that the IRS can only adjust the phase shift, i.e., $\left| {{\theta _{m,n}}} \right| = 1$), ${{\bf{G}}_{l,\mathcal{M}}} = [{{\bf{G}}_{l,1}},...,{{\bf{G}}_{l,M}}]$, ${{\bf{H}}_{R,\mathcal{M}}} = {[{{\bf{H}}^{T}_{R,1}},...,{{\bf{H}}^{T}_{R,M}}]^T}$,  $\mathbf{\Theta}  = \mathrm{diag}({\{ {\mathbf{\Theta} _m}\} _{m \in \mathcal{M}}})$, and ${{\bf{n}}_l} \sim \mathcal{CN}(\mathbf{0},{\sigma ^2}{\bf{I}})$ is the additive white Gaussian noise.

RRH \emph{l} compresses its received signals and then transmits the quantization bits to the BBU pool through a wired fronthaul link with limited capacity. By adopting the ¡®Gaussian test channel¡¯ model, the compressed signal recovered by the BBU pool can be expressed as \cite{Park14fronth}
\begin{equation}\label{compress_model}
{{\bf{\hat y}}_l} = {{\bf{y}}_l} + {{\bf{q}}_l},
\end{equation}
where ${\mathbf{q}_l} \sim \mathcal{CN}(0,{\mathbf{\Omega} _l})$ represents the quantization noise for RRH \emph{l} and ${\mathbf{\Omega} _l}$ denotes its covariance matrix which is determined by the corresponding quantization codebook.


\subsection{Uplink Sum Rate and Fronthaul Constraints}
From $(1)$ and $(2)$, the achievable uplink sum rate of the considered C-RAN is given by
\begin{equation}\label{up_sum_rate}
\begin{split}
&{R_{sum}}= I({\bf{x}};{{{\bf{\hat y}}}_{\mathcal{L}}})\\
&~~~~~~ = \log \left| P{{\bf{V}}_{\mathcal{L}}}{{\bf{V}}_{\mathcal{L}}^H}+{{\sigma ^2}{\bf{I}} + \mathbf{\Omega}_{\mathcal{L}} }  \right| - \log \left| {{\sigma ^2}{\bf{I}} + \mathbf{\Omega}_{\mathcal{L}} } \right|,
\end{split}
\end{equation}
where ${{\bf{\hat y}}_{\mathcal{L}}} = {[{\bf{\hat y}}_1^T,...,{\bf{\hat y}}_L^T]^T}$, ${\bf{V}_{\mathcal{L}}} = {\bf{H}_{\mathcal{L}}} + {\bf{G_{\mathcal{L}}\Theta }}{{\bf{H}}_{R,\mathcal{M}}}$, ${{\bf{H}}_{{\mathcal{L}}}} = [{{\bf{H}}_1^T},...,{{\bf{H}}_L^T}]^T$, $\mathbf{G_{\mathcal{L}}} = [{\mathbf{G}_{1,\mathcal{M}}^T},...,{\mathbf{G}_{L,\mathcal{M}}^T}]^T$ and ${\mathbf{\Omega}}_{\mathcal{L}}  = \mathrm{diag}(\{ {\mathbf{\Omega} _l}\} _{l \in \mathcal{L}})$.

The compression rate at each RRH should not exceed the fronthaul link capacity. With point-to-point compression, the corresponding fronthaul constraints are given by \cite{Park14fronth}
\begin{equation}\label{p2pconstraint_inf}
I({{\bf{y}}_{l}};{{\bf{\hat y}}_{l}}) \le {C_{l}},~\forall l\in\mathcal{L},
\end{equation}
where ${C_{l}}$ represents the fronthaul capacity from RRH $l$ to the BBU pool. According to \eqref{trans_model} and \eqref{compress_model}, the random variables in the mutual information term on the left hand side (LHS) follow joint Gaussian distribution. Let ${\bf{V}}_l = {\bf{H}}_l + {\bf G}_{l,\mathcal{M}}{\bf\Theta} {{\bf{H}}_{R,\mathcal{M}}}$. Constraint \eqref{p2pconstraint_inf} can be expressed as
\begin{equation}\label{ptpconstraint_cal}
\log \left| {P{{\bf{V}}_l}{\bf{V}}_l^H + {\sigma ^2}I + {\mathbf{\Omega}_l}} \right| - \log \left| {{\mathbf{\Omega}_l}} \right| \le {C_l}.
\end{equation}

Since the received signals at different RRHs are statistically dependent, RRHs can perform more efficient compression by exploiting this dependency via Wyner-Ziv coding. The BBU pool jointly recovers the compressed signals from all RRHs where the signal dependencies are utilized as side information.  The corresponding fronthaul constraints for Wyner-Ziv coding are given by \cite{Zhou2016CRANopt}
\begin{equation}\label{WZconstraint_inf}
I({{\bf{y}}_\mathcal{S}};{{\bf{\hat y}}_\mathcal{S}}|{{\bf{\hat y}}_{{\mathcal{\bar{S}}}}}) \le \sum\limits_{l \in \mathcal{S}} {{C_l}}, ~\forall \mathcal{S} \subseteq \mathcal{L},
\end{equation}
where ${\mathcal{\bar{S}}}$ is the complement set of ${\mathcal{S}}$. Similar to the case of point-to-point compression, one can evaluate the mutual information term on the LHS and rewrite constraint \eqref{WZconstraint_inf} as
\begin{equation}\label{WZconstraint_cal}
\begin{split}
&\log \left| P{{\bf{V}_{\mathcal{L}}}{\bf{V}_{\mathcal{L}}}^H + {\sigma ^2}{\bf{I}} + {\mathbf{\Omega}_{\mathcal{L}}}} \right| - \log \left| {{\mathbf{\Omega} _{\mathcal{S}}}} \right|\\
&- \log \left| P{{{\bf{V}}_{\mathcal{\bar{S}}}}{\bf{V}}_{\mathcal{\bar{S}}}^H + {\sigma ^2}{\bf{I}} + {\mathbf{\Omega} _{\mathcal{\bar{S}}}}} \right| \le \sum\limits_{l \in \mathcal{S}} {{C_l}},
\end{split}
\end{equation}
where ${{\bf{V}}_{\mathcal{\bar{S}}}} = {{\bf{H}}_{\mathcal{\bar{S}}}} + {{\bf G}_{\mathcal{\bar{S}}}\mathbf{\Theta} }{{\bf{H}}_{R,\mathcal{M}}}$ and ${\mathbf{\Omega}}_{\mathcal{\bar{S}}}  = \mathrm{diag}(\{ {\mathbf{\Omega} _l}\} _{l \in \mathcal{\bar{S}}})$.

\section{Joint Design of IRS Beamforming and Fronthaul Compression}
In this section, we investigate the joint design of IRS beamforming and fronthaul compression, aiming to maximize the system uplink sum rate. In the following, we first propose a joint optimization scheme for fronthaul compression with Wyner-Ziv coding. Then we extend the proposed scheme to the case of point-to-point fronthaul compression. Finally, we discuss the low complexity design under high SQNR regime.

\subsection{Sum Rate Maximization under Wyner-Ziv Coding}
With the goal of maximizing the uplink sum rate \eqref{up_sum_rate} under the fronthaul constraint \eqref{WZconstraint_cal} with Wyner-Ziv coding, we jointly optimize the IRS passive beamformers and the quantization noise covariance matrices for fronthaul compression. The problem is formulated as follows:
\[
\begin{split}
&\mathop {\max }\limits_{\mathbf{\Theta} ,\mathbf{\Omega}_{\mathcal{L}} }~\log \left| P{{{\bf{V}}_\mathcal{L}}{\bf{V}}_\mathcal{L}^H + {\sigma ^2}{\bf{I}} + {\bf{\Omega}}_\mathcal{L}} \right| - \log \left| {{\sigma ^2}{\bf{I}} + {\bf{\Omega }}_\mathcal{L}} \right|\\
&s.t.\log \left| P{{\bf{V}_{\mathcal{L}}}{\bf{V}_{\mathcal{L}}}^H + {\sigma ^2}{\bf{I}} + {\mathbf{\Omega}_{\mathcal{L}}}} \right| - \log \left| {{\mathbf{\Omega} _{\mathcal{S}}}} \right|\\
&~~~~~-\log \left| P{{{\bf{V}}_{\mathcal{\bar{S}}}}{\bf{V}}_{\mathcal{\bar{S}}}^H + {\sigma ^2}{\bf{I}} + {\mathbf{\Omega} _{\mathcal{\bar{S}}}}} \right| \le \sum\limits_{l \in \mathcal{S}} {{C_l}},~\forall \mathcal{S},\\
&~~~~~\left| {{\theta _{m,n}}} \right| = 1,\forall m,n,\\
\end{split}
\]
\begin{equation}\label{origin_P}
~~~~~~~~~{{\bf{\Omega }}_l} \succeq 0,\forall l.~~~~~~~~~~~~~~~~~~~~~~~~~~~~~~~~~~~~~~~~~~~~~
\end{equation}

It is non-trivial to find the optimal solution to the above problem, due to the fact that the objective function, the fronthaul constraints and the constraint $\left| {{\theta _{m,n}}} \right| = 1$ are not convex. In the following, we tackle the problem following the successive convex approximation approach \cite{Scutari2017Convexa}. Firstly, we approximate the non-convex objective and constraints to make the problem tractable.

Consider the objective in problem \eqref{origin_P}. By exploiting the Arimoto-Blabut algorithm\cite[Lemma 10.8.1, p. 33]{lem1}, we rewrite the objective \eqref{up_sum_rate} as follows:
\begin{equation}\label{lem_Obj}
{R_{sum}} = \mathop {\max }\limits_{q(\mathbf{x}|\mathbf{\hat y}_\mathcal{L})} \mathbb{E}\left[ {\log \frac{{q(\mathbf{x}|\mathbf{\hat y}_\mathcal{L})}}{{p(\mathbf{x})}}} \right],
\end{equation}
where the optimal ${q^*}(\mathbf{x}|\mathbf{\hat y}_\mathcal{L})$ for \eqref{lem_Obj} is the posterior probability $p(\mathbf{x}|\mathbf{\hat y}_\mathcal{L})$. According to \cite[Theorem 10.3, p. 32]{Them1}, $p(\mathbf{x}|\mathbf{\hat y}_\mathcal{L})$ follows the complex Gaussian distribution $\mathcal{CN}(\mathbf{W}^*\mathbf{\hat{y}_{\mathcal{L}}},\mathbf{\Sigma}^* )$ with
\begin{equation}\label{w_l}
{\mathbf{W}^ \ast }{\rm{ = }}\sqrt P {\mathbf{V}^H_\mathcal{L}}{\left( P{\mathbf{V}}_\mathcal{L}{\mathbf{V}}^H_\mathcal{L} + {\sigma ^2}\mathbf{I} + {\mathbf{\Omega}}_\mathcal{L}  \right)^{{\rm{ - 1}}}}
\end{equation}
\begin{equation}\label{sigma_l}
\mathrm{and}~{\mathbf{\Sigma} ^ \ast } = {\mathbf{I}} - \sqrt P{\mathbf{W}^* }{\mathbf V}_\mathcal{L}.
\end{equation}

Then we tackle constraint \eqref{WZconstraint_cal} in problem \eqref{origin_P}. Firstly we rewrite it as follows:
\begin{equation}\label{new1_C}
\begin{split}
&\log \left| {{{\bf{\Gamma }}_{\mathcal{L}}}} \right| - \log \left| {{{\bf{\Omega }}_\mathcal{S}}} \right| - \log \left| {{\sigma ^2}{\bf{I}} + {{\bf{\Omega }}_{{\mathcal{\bar{S}}}}}} \right|\\
&- {\log \left| {\frac{{{{\bf{V}}_{{\mathcal{\bar{S}}}}}{\bf{V}}_{{\mathcal{\bar{S}}}}^H + {\sigma ^2}{\bf{I}} + {{\bf{\Omega }}_{{\mathcal{\bar{S}}}}}}}{{{\sigma ^2}{\bf{I}} + {{\bf{\Omega }}_{{\mathcal{\bar{S}}}}}}}} \right|}\le \sum\limits_{l \in \mathcal{S}} {{C_l}},
\end{split}
\end{equation}
where ${{\bf{\Gamma }}_{\mathcal{L}}} = P{{\bf{V}}_{\mathcal{L}}}{\bf{V}}_{\mathcal{L}}^H + {\sigma ^2}{\bf{I}} + {{\bf{\Omega }}_{\mathcal{L}}}$. According to \cite{Zhou2016CRANopt}, the first term on the LHS is upper bounded by
\begin{equation}\label{first_1}
\log \left| {{{\bf{\Gamma }}_{\mathcal{L}}}} \right| \le \log \left| {{{\bf{{\rm \mathbf{E}}}}_{\mathcal{L}}}} \right| + \mathrm{Tr}({\bf{{\rm \mathbf{E}}}}_{\mathcal{L}}^{ - 1}{{\bf{\Gamma }}_{\mathcal{L}}}) - L{N_R},
\end{equation}
for ${\bf{{\rm \mathbf{E}}}}_{\mathcal{L}}\succeq 0$. The equality is achieved by
\begin{equation}\label{E_l}
{{\rm \mathbf{E}}_{\mathcal{L}}^ * } = \mathbf{\Gamma}_{\mathcal{L}}.
\end{equation}

The last log-term on the LHS of \eqref{new1_C} is equivalent to the mutual information $I\left({\bf{x}};{{{\bf{\hat y}}}_{\mathcal{\bar S}}}\right)$ under joint complex Gaussian distribution. Similar to the objective function, by exploiting the Arimoto-Blabut algorithm, we have
\begin{equation}\label{second}
\begin{split}
{\log \left| {\frac{{{{\bf{V}}_{{\mathcal{\bar{S}}}}}{\bf{V}}_{{\mathcal{\bar{S}}}}^H + {\sigma ^2}{\bf{I}} + {{\bf{\Omega }}_{{\mathcal{\bar{S}}}}}}}{{{\sigma ^2}{\bf{I}} + {{\bf{\Omega }}_{{\mathcal{\bar{S}}}}}}}} \right|}\ge {\rm \mathbb{E}}\left[ \log \frac{{\mathcal{CN}\left({{\bf{W}}_{{\mathcal{\bar{S}}}}}{{{\bf{\hat y}}}_{{\mathcal{\bar{S}}}}},{{\bf{\Sigma }}_{{\mathcal{\bar{S}}}}}\right)}}{{\mathcal{CN}(\mathbf{0},P{\bf{I}})}} \right],
\end{split}
\end{equation}
where the equality is achieve by
\begin{equation}\label{w_lc}
{\bf{W}}_{{\mathcal{\bar{S}}}}^ *{\rm{ = }}\sqrt P {\mathbf{V}^H_{\mathcal{\bar{S}}}}{\left( P{\mathbf{V}_{\mathcal{\bar{S}}}}{\mathbf{V}^H_{\mathcal{\bar{S}}}} + {\sigma ^2}\mathbf{I} + {\mathbf{\Omega}_{\mathcal{\bar{S}}}}  \right)^{{\rm{ - 1}}}}
\end{equation}
\begin{equation}\label{sigma_lc}
\mathrm{and}~
{\bf{\Sigma }}_{{\bar{S}}}^ * = {\mathbf{I}} - \sqrt P{{\bf{W}}_{{\mathcal{\bar{S}}}}^ *}{\mathbf{V}_{\mathcal{\bar{S}}}}.
\end{equation}

Therefore, we can approximate constraint \eqref{new1_C} as
\begin{equation}\label{new2_C}
\begin{split}
&\log \left| {{{\bf{E}}_{\mathcal{L}}}} \right| + \mathrm{Tr}({\bf{E}}_{\mathcal{L}}^{ - 1}{{\bf{\Gamma }}_{\mathcal{L}}})
- {\rm \mathbb{E}}\left[ \log (\frac{{\mathcal{CN}({{\bf{W}}_{{\mathcal{\bar{S}}}}}{{{\bf{\hat y}}}_{{\mathcal{\bar{S}}}}},{{\bf{\Sigma }}_{{\mathcal{\bar{S}}}}})}}{{\mathcal{CN}(0,P{\bf{I}})}}) \right]\\
& - \log \left| {{{\bf{\Omega }}_\mathcal{S}}} \right|- \log \left| {{\sigma ^2}{\bf{I}} + {{\bf{\Omega }}_{{\mathcal{\bar{S}}}}}} \right| - L{N_R} \le \sum\limits_{l \in \mathcal{S}} {{C_l}}
\end{split}
\end{equation}
With \eqref{lem_Obj} and \eqref{new2_C}, we reformulate the original problem \eqref{origin_P} as
\begin{equation}\label{new_P}
\begin{split}
&\mathop {\max }\limits_{\scriptstyle{\bf{W}},{\bf{\Sigma }},{\bf{\Theta }},{\bf{\Omega_\mathcal{L}}}\hfill\atop
\scriptstyle{{\bf{E}}_\mathcal{L}},{{\bf{W}}_{{\mathcal{\bar{S}}}}},{{\bf{\Sigma }}_{{\mathcal{\bar{S}}}}}\hfill} \mathbb{E}\left[ {\log \left( {\frac{{\mathcal{CN}({\bf{W}}{{{\bf{\hat y}}}_\mathcal{L}},{\bf{\Sigma }})}}{{\mathcal{CN}(0,P{\bf{I}})}}} \right)} \right]\\
&s.t.\eqref{new2_C},\forall \mathcal{S}\\
&~~~~\left| {{\theta _{m,n}}} \right| = 1,\forall m,n,\\
&~~~~{{\bf{\Omega }}_l} \succeq 0,\forall l.
\end{split}
\end{equation}
\emph{Remark 1}: According to \eqref{first_1} and \eqref{second}, any feasible solution to problem \eqref{new_P} is also feasible to the original problem \eqref{origin_P}, which indicates that we can solve problem \eqref{new_P} to obtain a sub-optimal solution to the original problem.

\subsection{Successive Approximation and Optimization}
In each iteration of the optimization algorithm, we first update the auxiliary variables ${{\bf{{\rm \mathbf{E}}}}_{\mathcal{L}}}$, ${\bf{W}}$, ${\bf{\Sigma }}$, ${\bf{W}}_{{\mathcal{\bar{S}}}} $ and ${\bf{\Sigma }}_{{\mathcal{\bar{S}}}}$, under fixed ${\bf{\Theta }}$ and ${\bf{\Omega_\mathcal{L}}}$. Obviously, the optimal ${\bf{W}}$ and ${\bf{\Sigma }}$ are given by \eqref{w_l} and \eqref{sigma_l}, respectively. On the other hand, ${{\bf{{\rm \mathbf{E}}}}_{\mathcal{L}}}$, ${\bf{W}}_{{\mathcal{\bar{S}}}}$ and ${\bf{\Sigma }}_{{\mathcal{\bar{S}}}}$ are updated according to \eqref{E_l}, \eqref{w_lc} and \eqref{sigma_lc}, respectively, which guarantees the convergence. We will discuss it later.

By fixing the auxiliary variables, we optimize $\mathbf{\Theta}$ and ${\mathbf{\Omega} _\mathcal{L}}$ in problem \eqref{new_P}, which can be regarded as an approximation of the original problem \eqref{origin_P}. 
Firstly, the expectation term in the objective can be evaluated as follows:
\[
\begin{split}
& ~~~- {\rm \mathbb{E}}\left[ {\log \left( {\frac{{\mathcal{CN}({\bf{W\hat y_{\mathcal{L}}}},{\bf{\Sigma }})}}{{\mathcal{CN}(0,P\mathbf{I})}}} \right)} \right]\\
&~{{ = \mathbb{E}[(}}{\bf{x}} - {\bf{W\hat  y_{\mathcal{L}}}}{{\rm{)}}^H}{{\bf{\Sigma }}^{ - 1}}{\rm{(}}{\bf{x}} - {\bf{W\hat  y_{\mathcal{L}}}}{\rm{)]}}{\rm{ + log}}\left| {\bf{\Sigma }} \right|{\rm{ - }}K\\
& \mathop = \limits^{(a)} {\hat{\theta} ^H}({\bf{A}} \odot {{\bf{B}}^T})\hat{\theta}  + 2{\mathop{\rm Re}\nolimits} (\mathrm{Tr}({\hat{\theta} ^H}{{\bf{z}}_1}))
+ \mathrm{Tr}({{\bf{W}}^H}{{\bf{\Sigma }}^{ - 1}}{\bf{W\Omega }}_\mathcal{L}) + {{{J}}_1}
\end{split}\]
\begin{equation}\label{new_o2}
\mathop = \limits^{(b)} \mathrm{Tr}({\bf{\Psi}\mathbf{\bar{\Theta}}}) + \mathrm{Tr}({{\bf{W}}^H}{{\bf{\Sigma }}^{ - 1}}{\bf{W\Omega }}) + {{{J}}_1},~~~~~~~~~~~~~~~~~~~~~~~~~~~~~~
\end{equation}
where in (a), we have the following notations: $\hat{\theta}  = \mathrm{diag}({\bf{\Theta }}),$
\[
\begin{split}
\mathbf{A} &= {\mathbf{G}_{\mathcal{L}}^H}{\mathbf{W}^H}{\mathbf{\Sigma} ^{ - 1}}\mathbf{W}\mathbf{G}_{\mathcal{L}},
\mathbf{B} = P{\bf{H}}_{R,\mathcal{M}}\mathbf{H}_{R,\mathcal{M}}^H,\\
{{\bf{z}}_1} &= \mathrm{diag}(P{{\bf{G}}_{\mathcal{L}}^H}{{\bf{W}}^H}{{\bf{\Sigma }}^{ - 1}}{\bf{W}}{{\bf{H}}_\mathcal{L}}{\bf{H}}_{R,\mathcal{M}}^H - \sqrt P {{\bf{G}}_{\mathcal{L}}^H}{{\bf{W}}^H}{{\bf{\Sigma }}^{ - 1}}{{\bf{H}}_{R,\mathcal{M}}^H}),\\
{{{J}}_1} &= \mathrm{Tr}(P{{\bf{W}}^H}{{\bf{\Sigma }}^{ - 1}}{\bf{W}}{{\bf{H}}_\mathcal{L}}{\bf{H}}_\mathcal{L}^H){\rm{ - }}2{\mathop{\rm Re}\nolimits}(\mathrm{Tr}(\sqrt P{{\bf{\Sigma }}^{ - 1}}{\bf{W}}{{\bf{H}}_\mathcal{L}} ))\\
&~{\rm{ + }}\mathrm{Tr}({\sigma ^2}{{\bf{W}}^H}{{\bf{\Sigma }}^{ - 1}}{\bf{W}}) + \mathrm{Tr}({{\bf{\Sigma }}^{ - 1}})+ {\rm{log}}\left| {\bf{\Sigma }} \right|- K,
\end{split}
\]
and in (b), we have $\bar \theta  = \left[\mathbf{\hat{\theta}}^T, 1\right]^T$, $\mathbf{\bar{\Theta}}  = \bar \theta {\bar \theta ^H}$ and
\[\mathbf{\Psi}  = \left( {\begin{array}{*{20}{c}}
\begin{array}{l}
\mathbf{A} \odot {\mathbf{B}^T}\\
\mathbf{z}_1^H
\end{array}&\begin{array}{l}
{\mathbf{z}_1}\\
0
\end{array}
\end{array}} \right).\]

Similarly we can evaluate constraint \eqref{new2_C} in problem \eqref{new_P} as follows:
\begin{equation}\label{new3_C}
\begin{split}
&\mathrm{Tr}(\mathbf{\Upsilon}_{\mathcal{\bar{S}}} {\bf{\bar \Theta }}) + \mathrm{Tr}({{\bf{E}}_{\mathcal{L}}^{ - 1}}{{\bf{\Omega }}_{\mathcal{L}}}) + \mathrm{Tr}({\bf{W}}_{{\mathcal{\bar{S}}}}^H{\bf{\Sigma }}_{{\mathcal{\bar{S}}}}^{ - 1}{{\bf{W}}_{{\mathcal{\bar{S}}}}}{{\bf{\Omega }}_{{\mathcal{\bar{S}}}}})\\
& - \log |{\sigma ^2}{\bf{I}} + {{\bf{\Omega }}_{{\mathcal{\bar{S}}}}}| - \log |{{\bf{\Omega }}_\mathcal{S}}| + {J_{2,\mathcal{\bar{S}}}} \le \sum\limits_{l \in \mathcal{S}} {{C_l}}.
\end{split}
\end{equation}
where
${{\bf{A}}_{\mathcal{\bar{S}}}}={\bf{G}}_{\mathcal{L}}^H{\bf{E}}_{\mathcal{L}}^{ - 1}{{\bf{G}}_{\mathcal{L}}} + {\bf{G}}_{{\mathcal{\bar{S}}}}^H{\bf{W}}_{{\mathcal{\bar{S}}}}^H{\bf{\Sigma }}_{{\mathcal{\bar{S}}}}^{ - 1}{{\bf{W}}_{{\mathcal{\bar{S}}}}}{{\bf{G}}_{{\mathcal{\bar{S}}}}},$
\[\begin{split}
{{\bf{z}}_{2,{\bar{\mathcal{S}}}}}&= \mathrm{diag}(P{\bf{G}}_{\mathcal{L}}^H{{\bf{E}}^{ - 1}}{{\bf{H}}_{\mathcal{L}}}{\bf{H}}_{R,\mathcal{M}}^H- \sqrt P{\bf{G}}_{{\mathcal{\bar{S}}}}^H{\bf{W}}_{{\mathcal{\bar{S}}}}^H{\bf{\Sigma }}_{{\mathcal{\bar{S}}}}^{ - 1}{\bf{H}}_{R,\mathcal{M}}^H\\
&+ P{\bf{G}}_{{\bar{\mathcal{S}}}}^H{\bf{W}}_{{\bar{\mathcal{S}}}}^H{\bf{\Sigma }}_{{\bar{\mathcal{S}}}}^{ - 1}{{\bf{W}}_{{\mathcal{\bar{S}}}}}{{\bf{H}}_{{\bar{\mathcal{S}}}}}{\bf{H}}_{R,\mathcal{M}}^H),\\
{J_{2,{\bar{\mathcal{S}}}}} &= \log \left| {{{\bf{E}}_{\mathcal{L}}}} \right| + \mathrm{Tr}({\bf{E}}_{\mathcal{L}}^{ - 1}({\sigma ^2}{\bf{I}} + P{{\bf{H}}_{\mathcal{L}}}{\bf{H}}_{\mathcal{L}}^H))- L{N_R}-K\\
&- 2{\mathop{\rm Re}\nolimits} \{ \mathrm{Tr}(\sqrt P{\bf{H}}_{{\bar{\mathcal{S}}}}^H{\bf{W}}_{{\bar{\mathcal{S}}}}^H{\bf{\Sigma }}_{{\bar{\mathcal{S}}}}^{ - 1})\}+ \log |{{\bf{\Sigma }}_{\bar{\mathcal{S}}}}| + \mathrm{Tr}({\bf{\Sigma }}_{\bar{\mathcal{S}}}^{ - 1})\\
&+ \mathrm{Tr}({\bf{W}}_{{\mathcal{\bar{S}}}}^H{\bf{\Sigma }}_{{\mathcal{\bar{S}}}}^{ - 1}{{\bf{W}}_{{\bar{\mathcal{S}}}}}({\sigma ^2}{\bf{I}} + P{{\bf{H}}_{{\mathcal{\bar{S}}}}}{{\bf{H}}_{{\bar{\mathcal{S}}}}})),\\
{\mathbf{\Upsilon} _{\mathcal{\bar{S}}}} &= \left( {\begin{array}{*{20}{c}}
\begin{array}{l}
{{\bf{A}}_{\mathcal{\bar{S}}}} \odot {{\bf{B}}^T}\\
{\bf{z}}_{2,\mathcal{\bar{S}}}^H
\end{array}&\begin{array}{l}
{{\bf{z}}_{2,\mathcal{\bar{S}}}}\\
0
\end{array}
\end{array}} \right).
\end{split}
\]

Now the optimization problem \eqref{new_P} becomes:
\begin{equation}\label{opt_p}
\begin{split}
&\mathop {\min }\limits_{{\bf{\bar \Theta }},{{\bf{\Omega }}_\mathcal{L}}} \mathrm{Tr}({\bf{\Psi \bar \Theta }}) + \mathrm{Tr}({{\bf{W}}^H}{{\bf{\Sigma }}^{ - 1}}{\bf{W}}{{\bf{\Omega }}_{\mathcal{L}}}) + {J_1}\\
&s.t.~\mathrm{Tr}(\mathbf{\Upsilon}_{\bar{\mathcal{S}}} {\bf{\bar \Theta }}) + \mathrm{Tr}({{\bf{E}}^{ - 1}}{{\bf{\Omega }}_{\mathcal{L}}})\\
&~~~~~ + \mathrm{Tr}({\bf{W}}_{{\bar{\mathcal{S}}}}^H{\bf{\Sigma }}_{\bar{\mathcal{S}}}^{ - 1}{{\bf{W}}_{\bar{\mathcal{S}}}}{{\bf{\Omega }}_{\bar{\mathcal{S}}}}) - \log |{\sigma ^2}{\bf{I}} + {{\bf{\Omega }}_{\bar{\mathcal{S}}}}|\\
&~~~~~ - \log |{{\bf{\Omega }}_\mathcal{S}}| + {J_{2,\bar{\mathcal{S}}}} \le \sum\limits_{l \in \mathcal{S}} {{C_l}},~\forall \mathcal{S}\\
&~~~~~\mathrm{rank}({\bf{\bar \Theta }}) = 1,{\bf{\bar \Theta }} \succ 0,|{{{\bf{\bar \Theta }}}_{i,i}}| = 1,~\forall i\\
&~~~~~{{\bf{\Omega }}_l} \succ 0,~\forall l.\\
\end{split}
\end{equation}

We apply SDR by relaxing the rank-one constraint and the resulted problem becomes convex. Thus it can be effectively solved by standard convex optimization tools like CVX. Note that the obtained $\mathbf{\bar{\Theta}}$ may not be exactly rank-one in general. We apply the efficient randomization techniques given in \cite{Sidiropoulos2006Rand} to generate suboptimal candidates and choose the one achieving the minimal objective function.

To this end, we summarize the proposed joint fronthaul compression and IRS beamforming optimization algorithm:
\begin{algorithm}
\caption{} 
\begin{algorithmic}[1]
\State Initialize ${\bf{\Theta }},{{\bf{\Omega }}_\mathcal{L}}$ feasible for problem \eqref{origin_P}.
\State Update ${\bf{W}}$, ${\bf{\Sigma }}$, $\mathbf{E}_{\mathcal{L}}$, ${\bf{W}}_{{\mathcal{\bar{S}}}}$, ${\bf{\Sigma }}_{{\mathcal{\bar{S}}}}$, using \eqref{w_l}, \eqref{sigma_l}, \eqref{E_l}, \eqref{w_lc}, \eqref{sigma_lc}, respectively.
\State Solve problem \eqref{opt_p}. Update ${\bf{\Theta }},{{\bf{\Omega }}_\mathcal{L}}$ if the objective decreases.
\State Repeat Step 2-3, until convergence.
\end{algorithmic}
\end{algorithm}

\emph{Remark 2:} The convergence of Algorithm $1$ can be proved by verifying that Steps $2-3$ induce a non-decreasing sequence of objective values for problem \eqref{new_P}. Consider the $t$th iteration. In Step $2$, recalling \eqref{lem_Obj}, updating $\{{\bf{W}}^{\left(t\right)},{\bf{\Sigma }}^{\left(t\right)}\}$ by \eqref{w_l} and \eqref{sigma_l} results in a non-decreasing objective value. On the other hand, updating other auxiliary variables according to \eqref{E_l}, \eqref{w_lc} and \eqref{sigma_lc} does not affect the objective. Nevertheless, it guarantees that the solution in the last iteration, i.e., $\{{\bf{\Theta }}^{\left(t-1\right)},{{\bf{\Omega }}_\mathcal{L}}^{\left(t-1\right)}\}$, is still feasible for problem \eqref{opt_p} under the updated auxiliary variables. In Step 3, since SDR is applied, we solely obtain a suboptimal solution to problem \eqref{opt_p}. Therefore, we check whether the obtained $\{{\bf{\Theta}}^{\left(t\right)}, {{\bf{\Omega }}_\mathcal{L}}^{\left(t\right)}\}$ decreases the objective value compared with that achieved by the solution in the last iteration.

\emph{Remark 3:}
The computational complexity of Algorithm 1 is dominated by Step 3 which involves solving problem \eqref{opt_p}. Since problem \eqref{opt_p} is convex after SDR, it can be efficiently solved by the interior-point method with computational complexity in the problem size given by ${(M{N_I})^2} + LN_R^2$ \cite{Zhou2016CRANopt}. The overall complexity of Algorithm $1$ is given as the product of the number of iterations and the above complexity.
\subsection{Extension to the Case of Point-to-Point Compression}
The extension to the case of point-to-point compression is straightforward, by replacing constraint \eqref{WZconstraint_cal} in problem \eqref{origin_P} with constraint \eqref{ptpconstraint_cal} for point-to-point compression. To solve the optimization problem, we similarly tackle the non-convex objective as in \eqref{lem_Obj}. The fronthaul constraint \eqref{ptpconstraint_cal} can be approximated as a convex constraint by applying \eqref{first_1} to the first log-term on its LHS. Then we can still apply the successive approximation and SDR optimization as in Section III.B. The corresponding algorithm is similar to Algorithm 1. Due to the limited space, the details are not given here.
\subsection{Low Complexity Design under High SQNR}
In \cite{Zhou2016CRANopt}, it was shown that for C-RAN uplink, in high SQNR regime, the quantization noise covariance matrices are nearly-optimal when proportional to the received noise covariance matrices at RRHs, i.e., ${{\bf{\Omega }}_l^*} \approx \beta_l{\bf{I}}$ for RRH $l$, where $\beta_l>0$ is chosen to satisfy the fronthaul constraint. Although it was proved for point-to-point compression and Wyner-Ziv coding with sequential decompression, it is straightforward to prove it in the case of Wyner-Ziv coding with joint decompression considered in our work. Therefore, we can propose an efficient scheme under high SQNR for both Wyner-Ziv coding and point-to-point compression. Explicitly, we set ${{\bf{\Omega }}_l} = \beta_l{\bf{I}}$ and then jointly optimize ${\beta _l}$ and $\bf\Theta$, wherein the non-convex objective function and constraints are tackled in the same way as before. Since we optimize the scalar ${\beta _l}$ instead of the whole covariance matrix ${{\bf{\Omega }}_l}$, the computational complexity is reduced. Moreover, note that in practice such ${{\bf{\Omega }}_l}$ can be realized by low-complexity per-antenna signal quantization at each RRH.
\section{Numerical Results}
In this section we present numerical results to validate the effectiveness of the proposed algorithms. In the simulation scenario, 4 users are uniformly distributed within a circle centered at the origin with radius of $30$m, $2$ RRHs each equipped with 4 antennas are located at (-30m,90m) and (30m,90m), respectively. $2$ IRSs are deployed at (-40m,80m) and (40m,80m). The path loss is modeled as $k = \xi {d^{ - \alpha }}$, where \emph{d} is the link distance, $\alpha $ is the path loss exponent, and $\xi$ is set to -30dB. We model both the user-IRS link and IRS-RRH link as LoS channel. The path loss exponent $\alpha$ for user-RRH link, user-IRS link, IRS-RRH link is set to 3.6, 2.2 and 2.2, respectively. As for small-scale fading, we assume that the user-RRH link follows Rayleigh fading, and the user-IRS link and the IRS-RRH link follow Rician fading with a Rician factor of 10dB. The Gaussian noise variance is set to -89dBm.
\begin{figure}
  \centering
\centering
  \includegraphics[width=0.44\textwidth]{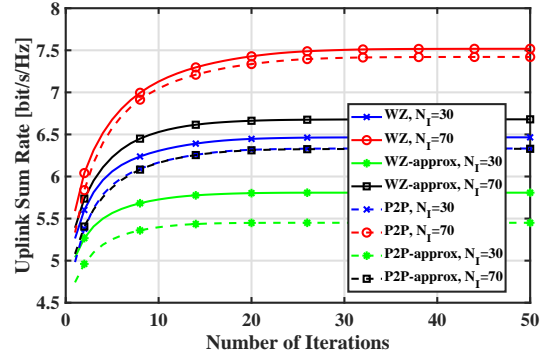}\\
  \caption{Average sum rate versus number of iterations.}
\end{figure}

Firstly, we verify the convergence of the proposed algorithms under different numbers of reflecting elements per IRS as shown in Fig. $1$, wherein `WZ' and `P2P' represent the proposed algorithms for Wyner-Ziv coding and point-to-point compression, respectively, and `WZ-approx' and `P2P-approx' denote the low-complexity scheme for high SQNR regime where we set ${{\bf{\Omega }}_l} = \beta_l{\bf{I}}$. It can be observed that the proposed algorithms converge in all cases.
\begin{figure}
  \centering
\centering
  \includegraphics[width=0.45\textwidth]{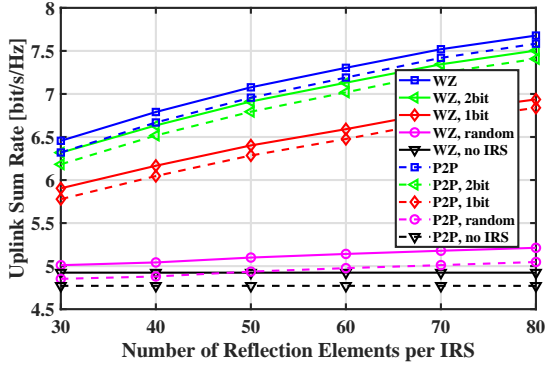}\\
  \caption{Average sum rate versus number of IRS elements, where $P$ = 10dBm, ${C_1}$ =$C_2$=5 bps/Hz.}
\end{figure}
\begin{figure}
  \centering
\centering
  \includegraphics[width=0.45\textwidth]{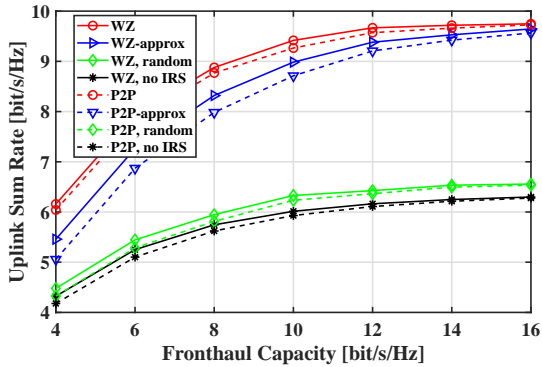}\\
  \caption{Average sum rate versus fronthaul capacity, where $P$ = 10dBm and ${N_I}$ = 50.}
\end{figure}

Fig. $2$ plots the average achieved uplink sum rate of the proposed algorithms versus the number of reflecting elements per IRS. For both Wyner-Ziv coding and point-to-point compression, we also simulate the following cases: 1)`$b$bit': in practice, the IRS phase shift cannot be continuous. It takes ${2^b}$ discrete value, i.e., ${\theta}_{m,n} = \{ 1,{e^{j\frac{{2\pi }}{{{2^b}}}}},...,{e^{j\frac{{({2^b} - 1)2\pi }}{{{2^b}}}}}\} $. In this case the optimized IRS phase shifts obtained by Algorithm 1 are projected to the nearest discrete values and ${{\bf{\Omega }}_l}$ is scaled to meet the fronthaul constraints; 2)`random': IRS phase shifts are uniformly distributed within $[0,2\pi )$ while only ${{\bf{\Omega }}_l}$ is optimized; 3)`no IRS': the IRS is removed while ${{\bf{\Omega }}_l}$ is optimized. In Fig. 2, it can be observed that for both Wyner-Ziv coding and point-to-point compression, the achieved sum rate increases along with the increase of ${N_I}$. Therefore, deploying IRS can enhance the system performance, especially with the proposed optimization algorithms. Furthermore, the restriction for discrete IRS phase shift (i.e., `2bit') solely brings limited rate loss. It is also shown that Wyner-Ziv coding generally outperforms point-to-point compression, in accordance with the existing literature.

Fig. 3 plots the average uplink sum rate achieved by the proposed algorithms for Wyner-Ziv coding and point-to-point compression as well as the low-complexity schemes for high SQNR. As the fronthaul capacity increases, the rate loss of `WZ-approx' and `P2P-approx' becomes smaller. This validates that setting the quantization noise covariance matrix proportional to the identity matrix is nearly optimal under high SQNR. It is also observed that the gain of Wyner-Ziv coding over point-to-point compression becomes ignorable under high fronthaul capacity, due to the fact that the quantization noise is considerably smaller than the received noise and does not dominate the uplink rate.

\section{Conclusions}
In this letter, we have studied a joint design of passive beamforming and fronthual compression for multi-IRS-aided C-RAN uplink. A successive approximation and optimization approach is proposed which efficiently optimizes the IRS beamformers and the quantization noise covariance matrices to maximize the uplink sum rate under point-to-point compression and Wyner-Ziv coding, respectively. Numerical results verify that deploying IRS can considerably improve the system uplink rate with the proposed optimization algorithm.

\end{document}